\documentclass[conference,onecolumn,12pt]{IEEEtran}
\usepackage{cite}
\usepackage{graphicx}

\begin{document}
\title{Oct-tree Method on GPU: \$42/Gflops Cosmological Simulation}
\author{\authorblockN{Naohito Nakasato}
\authorblockA{Department of Computer Science and Engineering\\
University of Aizu\\
Aizu-Wakamatsu, Fukushima 965-0815, Japan\\
Email: nakasato@u-aizu.ac.jp}
}

\maketitle

\begin{abstract}
The kd-tree is a fundamental tool in computer science.
Among others, an application of the kd-tree search
(oct-tree method) to fast evaluation of particle interactions
and neighbor search is highly important
since computational complexity of these problems
are reduced from $O(N^2)$ with a brute force method
to $O(N {\rm log} N)$ with the tree method where N is a number of particles.
In this paper, we present a parallel implementation of
the tree method running on a graphic processor unit (GPU).
We successfully run a simulation of structure formation
in the universe very efficiently.
On our system, which costs roughly \$900,
the run with $N \sim 2.87 \times 10^6$ particles
took 5.79 hours and executed $1.2 \times 10^{13}$ force evaluations in total.
We obtained the sustained computing speed of 21.8 Gflops
and the cost per Gflops of \$41.6/Gflops that is two and half times better than 
the previous record in 2006.
\end{abstract}

\IEEEpeerreviewmaketitle

\section{Introduction}
A gravitational many-body simulation technique
is fundamental in astrophysical simulations
because gravity force drives the structure formation in the universe.
Length scales arisen in the structure formation
range from less than 1 cm at aggregation of dust to more than $10^{24}$ cm 
at formation of cosmological structure.
In all scales, gravity is a key physical process to understand the structure formation.
The reason behind this is long-range nature of gravity.

Suppose we simulate the structure formation with $N$ particles, 
a flow of a many-body simulation is as follows.
First we calculate mutual gravity force between $N$ particles then integrate
orbits for $N$ particles and repeat this process as necessary.
Although it is simple, the force-calculation is a challenging task in regarding computational science.
A simple and exact method to do the force-calculation requires $O(N^2)$ computational complexity, 
which is prohibitively compute intensive with large $N$.
The exact force-calculation is necessary in some types of simulations such as
a few-body problems, numerical integration of planets orbiting around a star (e.g., the Solar system), 
and evolution of dense star clusters.
For simulations that do not require exact force, a several approximation techniques
have been proposed \cite{Hockney_1981, Barnes_1986, Greengard_1987}.
The particle-mesh/particle-particle-mesh method \cite{Hockney_1981} and the oct-tree method \cite{Barnes_1986}
reduce the computational complexity of the force-calculation to $O(N {\rm log} N)$.
The fast-multipole method (FMM) further reduces it to $O(N)$.
Among these methods, the oct-tree method has been used extensively 
in astrophysical simulations since its adaptive nature is essential to deal 
with clumpy structure in the universe (e.g., \cite{Bouchet_1988}).

Despite $O(N {\rm log} N)$ complexity, computational optimization
to the oct-tree method such as vectorization and parallelization is necessary
to accommodate demands for simulations with larger and large $N$.
In \cite{Hernquist_1990, Makino_1990, Barnes_1990}, they have reported 
various techniques to vectorize the force-calculation with the oct-tree method.
In \cite{Warren_1992, Dubinski_1996, Yahagi_1999}, they have reported
their parallel oct-tree method for massively parallel processors (MPP).
In a recent work \cite{Springel_2005}, with a parallel oct-tree code running on MPP, 
they have reported the simulation of large-scale structure formation in the universe
with more than ten billions particles.
Another computational technique that speed-up the oct-tree method is
to utilize special purpose computers GRAPE\cite{Sugimoto_1990, Makino_1998}.
A combination of vectorization techniques of the oct-tree method
and GRAPE, one can execute the oct-tree method efficiently on GRAPE \cite{Makino_1991} 

Notably, the cosmological simulation is a grand challenge problem.
In fact, the cosmological simulations were awarded many times in the Gordon Bell prizes
\cite{Warren_1992a, Fukushige_1996, Warren_1997, Warren_1998, Kawai_1999}.
In those work, both parallel tree codes \cite{Warren_1992a, Warren_1997, Warren_1998}
and a tree code with GRAPE \cite{Fukushige_1996, Kawai_1999} have
been adopted to do cosmological simulations.

In the present paper, we describe our implementation of the oct-tree method on 
a graphic processing unit (GPU).
The rise of the GPU forces us to re-think a way of parallel computing on it
since a performance of recent GPUs is impressive at $> 1$ Tflops.
Acceleration techniques for many-body simulations with GPU
have been already reported (\cite{Nyland_2007} and many others), 
however, they have implemented the exact but brute force method with $O(N^2)$ complexity.
Apparently, for applications that do not require the exact force, 
it is possible to do much efficient computation with the oct-tree method.
We have implemented the oct-tree method on GPU so that
we can enjoy the speed of $O(N {\rm log} N)$ algorithm on GPU.
With small $N < 10,000$, the brute force method on GPU 
is faster than the oct-tree method on GPU 
due to extra work concerning tree data structure.
However, our result show the oct-tree on GPU significantly 
outperform the brute force method with $N \gg 10,000$ at which
a standard size of $N$ in current astrophysical simulations is. 
As an application of our code, we present a result of a cosmological simulation
with our tree code running on RV770 GPU.

\section{Our Computing System with GPU}
Our computing system used in the present paper consists of a host computer and 
an extension board.
A main component of the extension board is a GPU processor that 
is acted as an accelerator attached to the host computer.
A program running of the host computer cooperates with a program running on the GPU to do useful tasks.

\subsection{Architecture of RV770 GPU}
In this section, we briefly summarize a GPU
that we used to implement the oct-tree method for cosmological simulations.

RV770 processor from AMD/ATi is the company's latest GPU (R700 architecture)
with many enhancements for general purpose computing on GPU (GPGPU).
It has 800 arithmetic units (called a stream core), each of which is capable of executing
single precision floating-point (FP) multiply-add in one cycle.
At the time of writing, the fastest RV770 processor is running at 750 MHz
and offers a peak performance of $800 \times 2 \times 750 \times 10^6 = 1.2$ Tflops.
Internally, there are two types of the stream cores in the processor.
One is a simple stream core that can execute only a FP multiply-add
and integer operations and operates on 32 bit registers. 
Another is a transcendental stream core that can handle 
transcendental functions in addition to the above simple operations.

Moreover, these units are organized hierarchically as follows.
At one level higher from the stream cores, a five-way very long instruction word
unit called a thread processor (TP), that consists of four simple stream cores
and one transcendental stream core.
Therefore, one RV770 processor has 160 TPs.
The TP can execute either at most five single-precision/integer operations, 
four simple single-precision/integer operations with one transcendental operation, 
or double-precision operations by combinations of the four stream cores.
Moreover, a unit called a SIMD engine consists of 16 TPs.
Each SIMD engine has a memory region called a local data store
that can be used to explicitly exchange data between TPs.

\begin{figure}
\centering
\includegraphics[width=13cm,angle=-90]{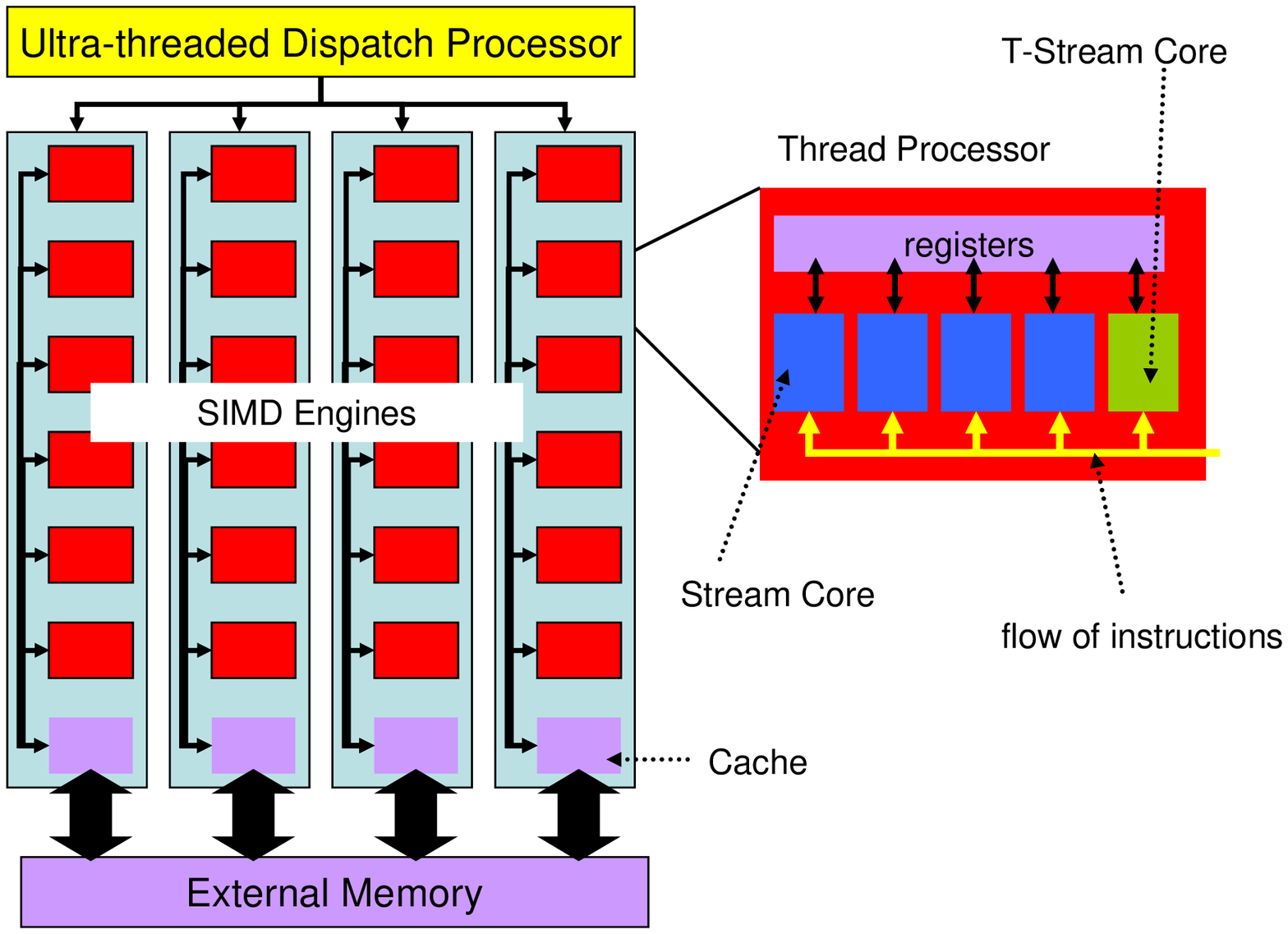}
\caption{A block diagram of RV770 GPU. Note a number of components is not exactly same
as the actual numbers}
\label{RV770}
\end{figure}

At the top level RV770, there are 10 SIMD engines, 
a controller unit called an ultra-threaded dispatch processor, 
and other units such as units for graphic processing, 
memory controllers and DMA engines.
An external memory attached to the RV770 in the present work
is 1 GB GDDR5 memory with a bus width of 256 bit.
It has a data clock rate at 3600 MHz and offers us a bandwidth of 115.2 GB sec$^{-1}$.
In addition to this large memory bandwidth, 
each SIMD engine on RV770 has two-level cache memory.
Figure \ref{RV770} shows a block diagram of RV770.

The RV770 processor with memory chips is mounted on an extension board.
The extension board is connected with a host computer through PCI-Express Gen2 x16 bus.
A theoretical communication speed between the host computer and RV770 GPU is
at most 8 GB sec$^{-1}$ (in one-way).
The measured communication speed of our system (shown in Table \ref{conf})
is $\sim 5 - 6$ GB sec$^{-1}$ for data size larger than 1 MB.

\subsection{Programming RV770 GPU}
In 2008, AMD/ATi has released a software development kit (SDK) for 
stream computing/GPGPU on their GPU products.
The SDK consists of two levels as a compute abstraction layer (CAL)
and a high-level language (called Brook+) similar to the C Language.
The CAL is more or less close to a bare hardware and offers us
a basic library that enables us to access, manage and control GPU(s)
and to generate machine instructions from an assembly like language
called IL (intermediate language).
The IL is like a virtual instruction set for GPU from AMD/ATi.
In the present work, we program the RV770 GPU using the IL
to gain full control of RV770 GPU. 

A programming model supported by CAL and IL is
a single instruction and multiple data (SIMD) at the level of TP.
In this programming model, a sequence of instructions generated
from an IL program is executed on all TPs simultaneously
with different input data from a viewpoint of a user.
Internally, the ultra-threaded dispatch processor
controls a flow of processing.
It can supply different instructions to different SIMD engines
so that TPs in a SIMD engine executes the same instructions
but other SIMD engines can execute different instructions.
Therefore, actual execution of a program 
on RV770 is like a multiple program and multiple data (MPMD) model.

A code written in IL is called a compute kernel.
In a compute kernel, we explicitly declare which type of variable input data is.
In a main-body of the IL code, we write arithmetic operations on input data.
Logically, each TP is implicitly assigned data that is different each other. 
In a simple compute kernel, it operates on the assigned data.
An operation like this, such as a pure stream computing, 
seems to work in highest efficiency.
In a complex compute kernel, which we explore in the present work, 
each TP not only operates on the assigned data
but also explicitly load random data that might be assigned to another TP.
To accomplish a random access to external memory, 
we explicitly calculate an address of data in our compute kernel.

\subsection{A Bare Many-body Performance on RV770 GPU}
So far, we have developed a several IL codes that can be used
to do astrophysical many-body simulations.
In this section, we report a performance of our implementation of 
the brute force method for computing gravity.
This code served as fundamental for us to implement an algorithm that is more sophisticated later.

Precisely, we have implemented conventional equations expressed as
\begin{eqnarray}
p_i &=& \sum_{j=1,j \ne i}^{N} p(\mbox{\boldmath{$x$}}_i, \mbox{\boldmath{$x$}}_j, m_j) =  
         \sum_{j=1,j \ne i}^{N} \frac{m_j}
                {(|\mbox{\boldmath{$x$}}_i - \mbox{\boldmath{$x$}}_j|^2 + \epsilon^2)^{1/2}}, \nonumber \\
\mbox{\boldmath{$f$}}_{i} &=& \sum_{j=1,j \ne i}^{N} \mbox{\boldmath{$f$}}(\mbox{\boldmath{$x$}}_i, \mbox{\boldmath{$x$}}_j, m_j) = \sum_{j=1,j \ne i}^{N} \frac{m_j (\mbox{\boldmath{$x$}}_i - \mbox{\boldmath{$x$}}_j)}
                {(|\mbox{\boldmath{$x$}}_i - \mbox{\boldmath{$x$}}_j|^2 + \epsilon^2)^{3/2}}, \nonumber \\
\label{gravity}
\end{eqnarray} 
where $\mbox{\boldmath{$f$}}_{i}$ and $p_i$ are force vector and potential
for a particle $i$, and $\mbox{\boldmath{$x$}}_i$, $m_i$, $\epsilon$ are 
position of a particle, the mass, and a parameter that prevents division by zero, respectively.
We can calculate these equations as a two-nested loop on a general purpose CPU.
In the most inner loop, by simultaneously evaluating functions $p$ and $\mbox{\boldmath{$f$}}$, 
we require 22 arithmetic operations, which include one square root and one division, 
to compute an interaction between particle $i$ and $j$.
Since previous authors starting from \cite{Warren_1997} used a conventional operational
count for evaluation of $\mbox{\boldmath{$f$}}_{i}$ and $p_i$, 
we also adopt the conventional counts of 38 throughout the paper. 

We simply implemented the summations Eq.(\ref{gravity}) on RV770
and have obtained a performance of $\sim 300$ GFLOPS for $N = 100,000$.
Since the peak performance of RV770 is more than four times larger, 
we have tried to optimize the simple implementation to get full utilization of the system.
In \cite{Elsen_2007}, they have reported their implementation of 
the brute force method for gravity and other forces on an older GPU from AMD/ATi.
A main insight they have obtained was that 
a loop unrolling technique greatly enhanced the performance of their code.
We have followed their approach and tried a several different ways of the loop unrolling.
Details of optimization will be presented elsewhere.

\begin{figure}
\centering
\includegraphics[width=12cm,angle=-90]{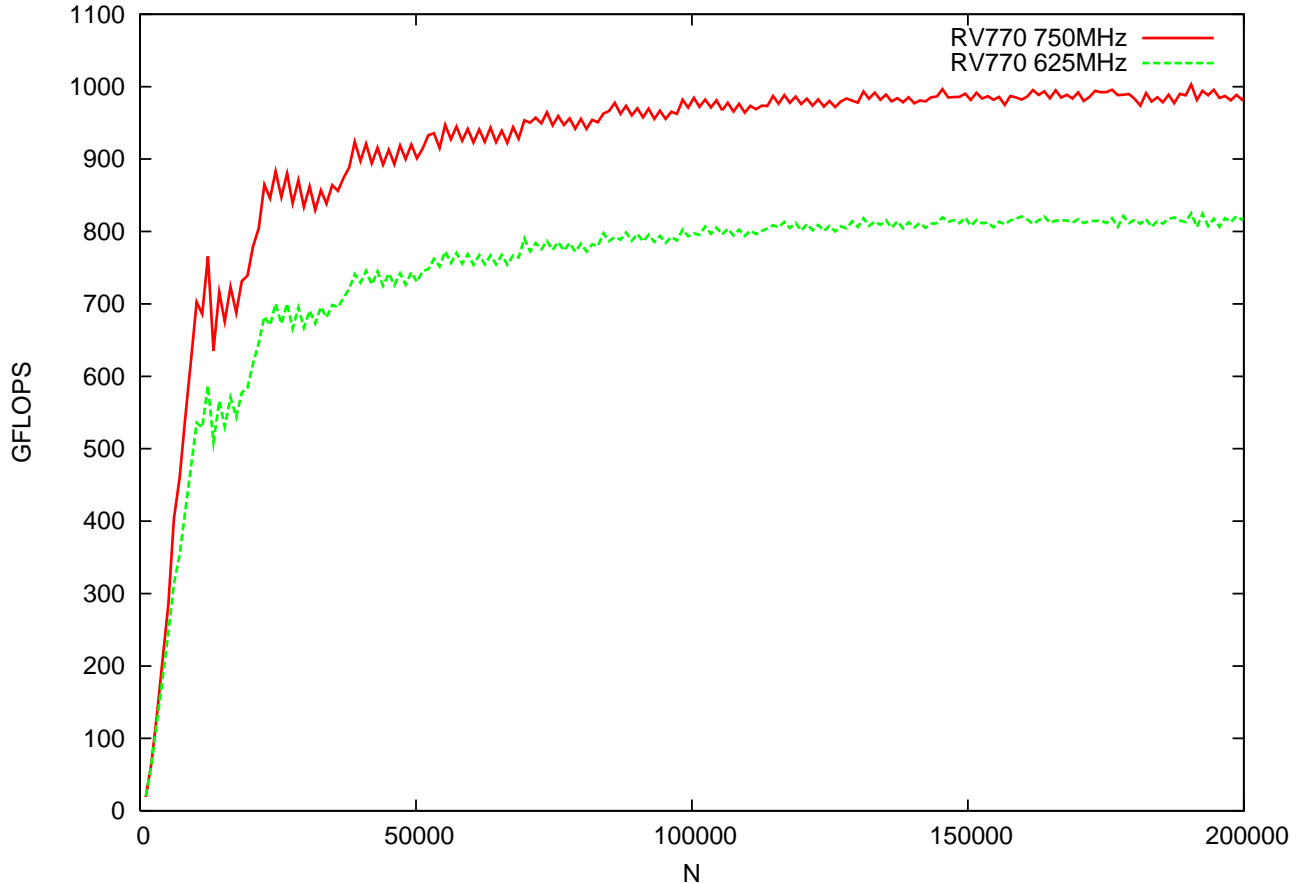}
\caption{A performance of the brute force method on two types of RV770 GPU}
\label{bare}
\end{figure}

In Figure \ref{bare}, we plot a computing speed of our optimized IL code for computing Eq.(\ref{gravity})
as a function of $N$. 
We have tested two configurations as one RV770 GPU running at 625 and 750 MHz, respectively.
So far, we have obtained a maximum performance of $\sim 990$ GFLOPS with $N \sim 200,000$. 
With $N = 226,304$, our optimized brute force method took roughly 2 seconds on one RV770 running at 750MHz.
As far as we know, the performance we obtained is fastest ever with one GPU chip.
Even with massive computing force available on GPU, 
we can not escape from a computational complexity of $O(N^2)$.
Therefore, if we need to do an astrophysical many-body simulation with large $N$, 
we need a smart algorithm to do the job
provided that recent standard of $N$ in astrophysical simulations
is at least $100,000$ for a complex simulation with baryon physics
and $1,000,000$ for a simple many-body simulation.

\section{Oct-tree Method on GPU}

\subsection{Oct-tree Method}
The oct-tree method \cite{Barnes_1986}
is a special case of the general kd-tree algorithm.
This method is optimized to efficiently calculate mutual force between particles 
and reduce computational complexity of the force-calculation 
from $O(N^2)$ with the brute force method to $O(N {\rm log} N)$. 
A trick is that instead of computing exact force with the brute force method, 
it approximates the force from distant particles
with multipole expansion generated by those distant particles.
Apparently, there is a trade-off between approximation error
and a way in which we replace a group of distant particles with their multipole expansion.
A tree structure that contains all particles
is used to efficiently judge this trade-off as we briefly explained.

The force-calculation with the oct-tree method is executed in two steps. 
(1) tree construction and (2) the force-calculation.
In the tree construction, we equally divide a cube that encloses whole particles
into eight sub-cells.
This first cell is a root of a tree that we construct.
It is called a root cell.
Then, each sub-cell is recursively sub-divided in the same say
until a cell has zero or one particle.
As the result of this procedure, we obtain the oct-tree.

\begin{figure}
\centering
\begin{verbatim}
procedure treewalk(i, cell)
  if cell has only one particle
    force += f(i, cell)
  else 
    if cell is far enough from i 
      force += f_multipole(i, cell)
    else 
      for i = 0, 7
        if cell->subcell[i] exists
          treewalk(i, cell->subcell[i])
\end{verbatim}
\caption{A pseudo code for the force-calculation by traversing the oct-tree}
\label{treewalk}
\end{figure}

In the force-calculation, we traverse the oct-tree
to judge whether we replace a distant cell, 
which has a group of particles that are geometrically close, 
with their multipole expansion.
If we do not replace, we further traverse sub-cells of the distant cell.
If we do replace, we calculate particle-cell interaction.
When we encounter a particle, we immediately calculate particle-particle interaction.
Given a particle (expressed as index \verb|i|) on which we want to compute the force acting,
this procedure is expressed as a pseudo code in Figure \ref{treewalk}.
Note \verb|subcell[]| are pointers to own sub-cells.
In this pseudo code, \verb|f| is a function that 
computes particle-particle interaction.
In addition \verb|f_multipole| is a function that computes particle-cell interaction.
In the present work, since we only consider monopole moment of a cell,
both functions are exactly expressed as Eq.(\ref{gravity}).
In principle, we can use any high order moment in particle-cell interaction.

We use this procedure starting from the root cell
with the following condition that tests whether a cell is far enough.
Let the distance between the particle and the cell is $d$.
The cell is well separated from the particle
if $l/d < \theta$, where $l$ is the size of the cell
and $\theta$ is a parameter that controls the trade-off.
Since smaller $l/d$ the cell is more distant from the particle, 
this condition (called opening condition) geometrically tests
whether the cell is far from the particle.
This recursive force-calculation procedure is almost same as the original algorithm
by \cite{Barnes_1986}.

An important nature of the oct-tree method is that
the force-calculation with tree traversal for different particles
is completely independent each other.
Therefore, after we complete the tree construction, 
the force-calculation is a massively parallel problem.
To take an advantage of this nature, there are two possibilities
to implement the oct-tree method on GPU.

\subsection{Oct-tree Method with GRAPE}
One is a method proposed by \cite{Makino_1991}.
This method has been proposed as a tree method for a special purpose computer GRAPE.
A system with GRAPE consists of a host computer and GRAPE board(s).
The host computer controls the GRAPE.
For a program running on the host, the GRAPE acts like
a subroutine that calculates gravity with given particles.

So we need the following two steps to use GRAPE for 
the force-calculation using the tree.
(1) construction of interaction list on the host computer
and (2) actual force-calculation on the GRAPE.
The interaction list is a list of particles and distant cells
that are supposed to interact with a given particle.
After construction of interaction lists for each particles are completed, 
we compute the force for each particles with GRAPE
by sending interaction lists to GRAPE.
These two steps are necessary because the GRAPE does not have
ability to traverse the tree.
Many authors have extensively used this method.
Two winners and a finalist of Gorden-Bell have used a variant of this method
with different version of GRAPE \cite{Fukushige_1996, Kawai_1999, Kawai_2006}.
In principle, we can adopt this method for GPU.
Actually, many authors have somehow emulated GRAPE with GPUs
(e.g., \cite{Nyland_2007, Hamada_2007, Belleman_2008})
and they all have obtained a good performance.
A drawback of this approach is that the performance is limited by
a speed of a host computer that is responsible for the tree traversal.
This possible bottleneck similar to the Amdahl's law might be critical without 
highly tuned \verb|treewalk()| implementation running on the host.
Furthermore, in all results by \cite{Fukushige_1996, Kawai_1999, Kawai_2006}, 
they have required a factor of two extra force evaluations
to obtain their best performance.
Note because of extra force evaluations, they have reported that
maximum error in force was better than the error obtained by 
the conventional oct-tree method with given $\theta$.

\subsection{General Treewalk}
Another way that we have taken in the present work is to implement the whole procedure
shown in Figure \ref{treewalk} on GPU.
Advantage of our approach is that only the tree construction, 
which requires relatively little time, is executed on a host so
that we utilize massive computing power of GPU as much as possible.
More importantly, we can apply our method to implement 
an application that requires short-range force interaction \cite{Warren_1995}.
This is because it is possible to implement neighbor search algorithm
as a general treewalk procedure shown in Figure \ref{general_treewalk}.
Two procedures \verb|proc_particle| and \verb|proc_cell|
are used to process particle-particle and particle-cell interaction, respectively. 
In addition, a function \verb|distance_test| is used to control the treatment of a distant cell.
The gravity force-calculation is an application of 
the general treewalk procedure that is very successful.

\begin{figure}
\centering
\begin{verbatim}
procedure general_treewalk(i, cell)
  if cell has only one particle
    proc_particle(i, cell)
  else 
    if distance_test(i, cell) is true
      proc_cell(i, cell)
    else 
      for i = 0, 7
        if cell->subcell[i] exists
          general_treewalk(i, cell->subcell[i])
\end{verbatim}
\caption{A pseudo code for a general treewalk procedure.} 
\label{general_treewalk}
\end{figure}

\subsection{Our GPU implementation}
In our implementation of the oct-tree method on GPU, we first construct
an oct-tree on a host computer that controls RV770.
At this stage, there is no difference between our original tree code
and a newly developed code for RV770.

\begin{figure}
\centering
\includegraphics[width=13cm]{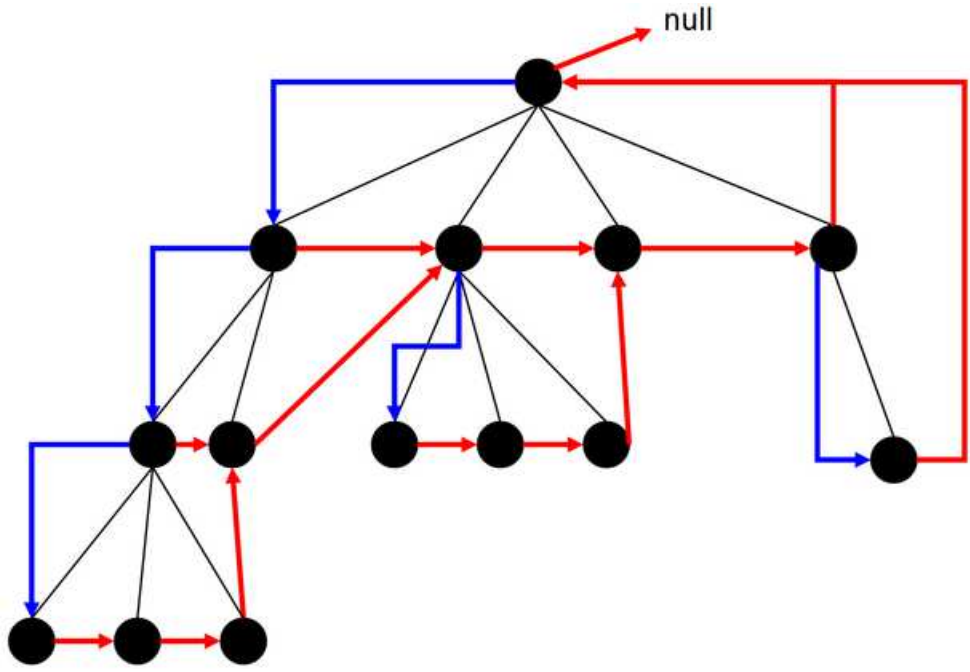}
\caption{A tree with more and next pointers,
which are shown in blue and red arrows, respectively.}
\label{tree}
\end{figure}

We need a special care to implement the treewalk procedure on RV770.
Currently, the IL does not support a recursive procedure 
except when it is possible to fully expand a recursion.
Such full expansion is only possible if a level of the recursion is definite
but in the tree method, we can't know how deep the recursion without a tree traversal is.
So we adopt a method proposed by \cite{Makino_1990}
that transform a recursion in \verb|treewalk()| into an iteration.
A key is that for a given cell
we does not need whole pointers (\verb|subcell[]|) to traverse the tree.
We only need two pointers to cells that we will visit next
when the opening condition is true and false, respectively.
These two pointers (hereafter we call \verb|next[]| and \verb|more[]|)
are easily obtained by a breath-first traversal on the tree.
Figure \ref{tree} shows \verb|next[]| and \verb|more[]| schematically.
Note a cell that has sub-cells has both \verb|next[]| and \verb|more[]| pointers
while a leaf cell (a particle in the present case) with no sub-cell has only \verb|next[]| pointer.
With the two pointers, an iterative form of \verb|treewalk()| is 
shown in Figure \ref{treewalk_iterative}.

\begin{figure}
\centering
\begin{verbatim}
procedure treewalk_iterative(i)
  cell = the root cell
  while cell is not null
    if cell has only one particle
      force += f(i, cell)
      cell = cell->next
    else 
      if cell is far enough from i 
        force += f_multipole(i, cell)
        cell = cell->next
      else 
        cell = cell->more
\end{verbatim}
\caption{A pseudo code for an iterative treewalk procedure.}
\label{treewalk_iterative}
\end{figure}

\begin{figure}
\centering
\includegraphics[width=13cm, angle=-90]{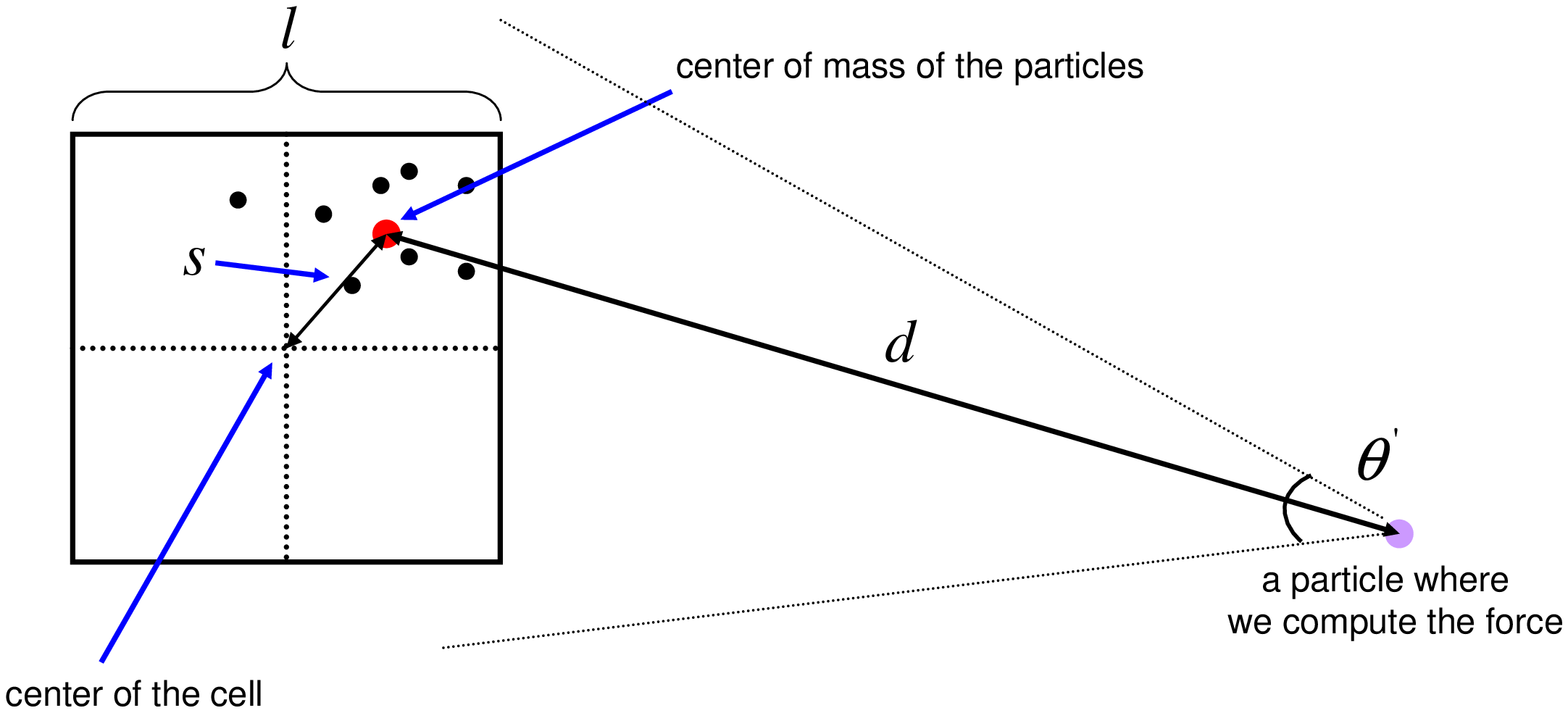}
\caption{A schematic view of a distant cell and a particle (shown in a solid purple point).
Black solid points are particles that belong to the cell.
A large red point is the center of the mass of the particles in the cell.
}
\label{cell-theta}
\end{figure}

The IL allow us to implement the iterative \verb|treewalk()| rather directly.
Input data for this compute kernel is four arrays.
First is a position and mass of particles and cell.
We pack a position and mass into a vector variable with four-element.
Therefore, it is the array of four-element vectors.
The mass of the cell equals to the total mass of particles in the cell.
And the position of the cell is at the center of mass of the particles.
Second and third arrays are the next and more pointers, respectively.
Both of them are a simple array.
And fourth array is the size of the cells.
The size of the cell is necessary for testing the opening condition.
In the present work, we adopt a following modified opening condition expressed as
\begin{equation}
\frac{l}{\theta} + s < d,
\label{mac}
\end{equation}
where $s$ is a distance between the center of a cell and
the center of mass of the cell \cite{Barnes_1994}.
The modified condition Eq.(\ref{mac}) takes into account 
particle distribution in a cell through $s$ since 
if particles gathers at a corner of a cell, 
the effective size of the cell become larger.
In Figure \ref{cell-theta}, we present a schematic view of a distant cell and a particle
on which we are trying to calculation force acting.
Practically, we pre-compute the square of effective size $S_{\rm effective}$ as
\begin{equation}
S_{\rm effective} = \left( \frac{l}{\theta} + s \right)^2,
\end{equation}
and send $S_{\rm effective}$ instead of $l$ for each cell.
With $S_{\rm effective}$, we don't need to computer square root of $d$
so that we simply compare $S_{\rm effective}$ and
$d^2$ during the tree traversal.

In Figure \ref{treegpu}, we present an abstracted version of
our compute kernel written in IL.
In the computing model of CAL, each TP execute the compute kernel
with assigned data in parallel.
In this code, \verb|own| represents the specific cell assigned to each TP.
\verb|=|, \verb|load| and \verb|->| and are not a real IL instruction or operation
but conventional idioms used here for explanation.
We omit a calculation of load address for arrays since it is somehow too detailed.
In addition the particle-particle and particle-cell interaction codes are omitted
because they simply compute functions $\mbox{\boldmath{$f$}}$ and $p$ in Eq.(\ref{gravity}).
As far as we understood, programming in IL is as much like 
programming an inner-most loop for a vector processor in an assembly language.
In this analogy, a group of TPs corresponds to a vector-processing unit.

\begin{figure}
\centering
\begin{verbatim}
...declaration of I/O arrays and constants...
...initialize variables for accumulation...

xi = load own->x
yi = load own->y
zi = load own->z

cell = root
whileloop
  break if cell is null

  xj = load cell->x
  yj = load cell->y
  zj = load cell->z
  mj = load cell->m
  s_eff = load cell->s_eff

  dx = xj - xi
  dy = yj - yi
  dz = zj - zi
  r2 = dx*dx + dy*dy + dz*dz

  if cell is a particle
    ...compute particle-particle interaction...
    cell = load next
  else 
    if r2 > s_eff
      ...compute particle-cell interaction...
      cell = load cell->next
    else 
      cell = load cell->more
  endif
endloop
\end{verbatim}
\caption{An abstracted IL code for our compute kernel
that executes the iterative treewalk.}
\label{treegpu}
\end{figure}

With the compute kernel we have shown, a flow of our oct-tree method on RV770 GPU is as follows. 
\begin{enumerate}
\item construct a tree ({\bf host})
\item compute the total mass, the center of mass, the effective size of each cell ({\bf host})
\item compute the next and more pointers ({\bf host})
\item send input data to GPU ({\bf host})
\item iterative treewalk associated with the force-calculation for each particle ({\bf GPU}) 
\item receive the force for each particle from GPU ({\bf host})
\end{enumerate}
We indicate whether the corresponding part is executed on either the host or RV770 GPU
with the bold text at the end of each step.

\subsection{Test and Optimization}
\label{btest}

\begin{table}
\renewcommand{\arraystretch}{1.3}
\caption{The configuration and cost of our system}
\label{conf}
\begin{center}
\begin{tabular}{c|c|c}
\hline
& Part  & price (JYE) \\
\hline
CPU & Core2 E8400 & 17,000 \\
MB  & ASUS P5E WS    & 27,900 \\
Memory & DDR2 800 1GB x4 & 4,000  \\
HDD & SATA 120GB         & 4,000  \\
Power unit & Schythe CorePower3 600 W & 7,780 \\
GPU & HD4780 memory 1GB & 30,000 \\ 
\hline
total price & & 90,680 \\
\hline
\end{tabular}
\end{center}
\end{table}

Here, we describe results of basic tests to see our code works correctly
and performance characteristics.
We used a configuration shown in Table \ref{conf} for all results presented in this paper.
The configuration is not a newest one but this setup shows best performance
among a several configurations we have tested.
In the basic tests, we used a randomly distributed particle in a sphere.

First, we see how a computing time depends on $N$ as shown in Table \ref{bench1}.
Each computing time were obtained by averaging results of 20 runs.
In this test, we set $\theta = 0.6$.
T$_{\rm total}$ and T$_{\rm construction}$ are the total time required
for the force-calculation and the time spent on the construction of tree, respectively.
Roughly, the tree construction took 20 - 28 \% of T$_{\rm total}$.
Note in all $N$, we have checked effectively there is no error in the force computed by RV770 GPU.
All operations on GPU were done with single precision so that we observed
that the error was comparable to the machine epsilon $\sim 10^{-6}$.
An exact cause of error is not fully clear, but we speculate the error is originated
from a difference in implementation of inverse of square root on host and GPU.
We think this is not significant at all for our purpose of astrophysical many-body simulations.

\begin{table}
\renewcommand{\arraystretch}{1.3}
\caption{Dependence of a computing speed on $N$ : no sorting}
\label{bench1}
\begin{center}
\begin{tabular}{c|c|c}
\hline
$N$ & T$_{\rm total}$ & T$_{\rm construction}$ \\
\hline
50K  & $4.70 \times 10^{-2}$ & $9.9 \times 10^{-3}$ \\
100K & $1.10 \times 10^{-1}$ & $2.1 \times 10^{-2}$ \\
200K & $2.69 \times 10^{-1}$ & $5.9 \times 10^{-2}$ \\
400K & $6.62 \times 10^{-1}$ & $1.7 \times 10^{-1}$ \\
800K & $1.53 \times 10^{0}$  & $4.3 \times 10^{-1}$ \\
\hline
\end{tabular}
\end{center}
\end{table}

Regarding a computing speed, randomly distributed particles are most severe because
two successive particles in input data are at different position in very high chance.
By its nature of the tree method, if two particles are close each other, 
those particles are expected to be at a same cell and interacts with a similar list
of particles and distant cells.
This means that if two successive particles in input data are geometrically close,
the treewalk of the second particle almost certainly takes less time due to a higher cache-hit rate.
To accomplish such situation, we can sort the particles to satisfy
successive particles are as close as possible.
Fortunately, such sorting is easily available with the tree method
by traversing tree with depth-first order.
In the course of the traversal, we add a particle encountered at each leaf node to a list.
After the tree traversal, we can use the obtained list to shuffle the particles
so that order of particles is in nearly desirable.
This ordering of particles is called the Morton ordering.
With this pre-processing, the speed of our method was altered as shown in Table \ref{bench2}.
Note the time in Table \ref{bench2} does not contain the time required for the pre-processing.
This is adequate since in astrophysical many-body simulations, tree is repeatedly 
constructed in each timestep so that we can automatically obtain this sorting for free.
Depending on $N$, we observed T$_{\rm total}$ obtained with the Morton ordering
is faster by a factor of 1.5 - 2.2 than without the pre-processing.
Moreover, T$_{\rm construction}$ also decreased in all cases due to better cache usage on the host.
With the Morton ordering, the tree construction took  roughly 14 - 27 \% of T$_{\rm total}$.

\begin{figure}
\centering
\includegraphics[width=12cm, angle=-90]{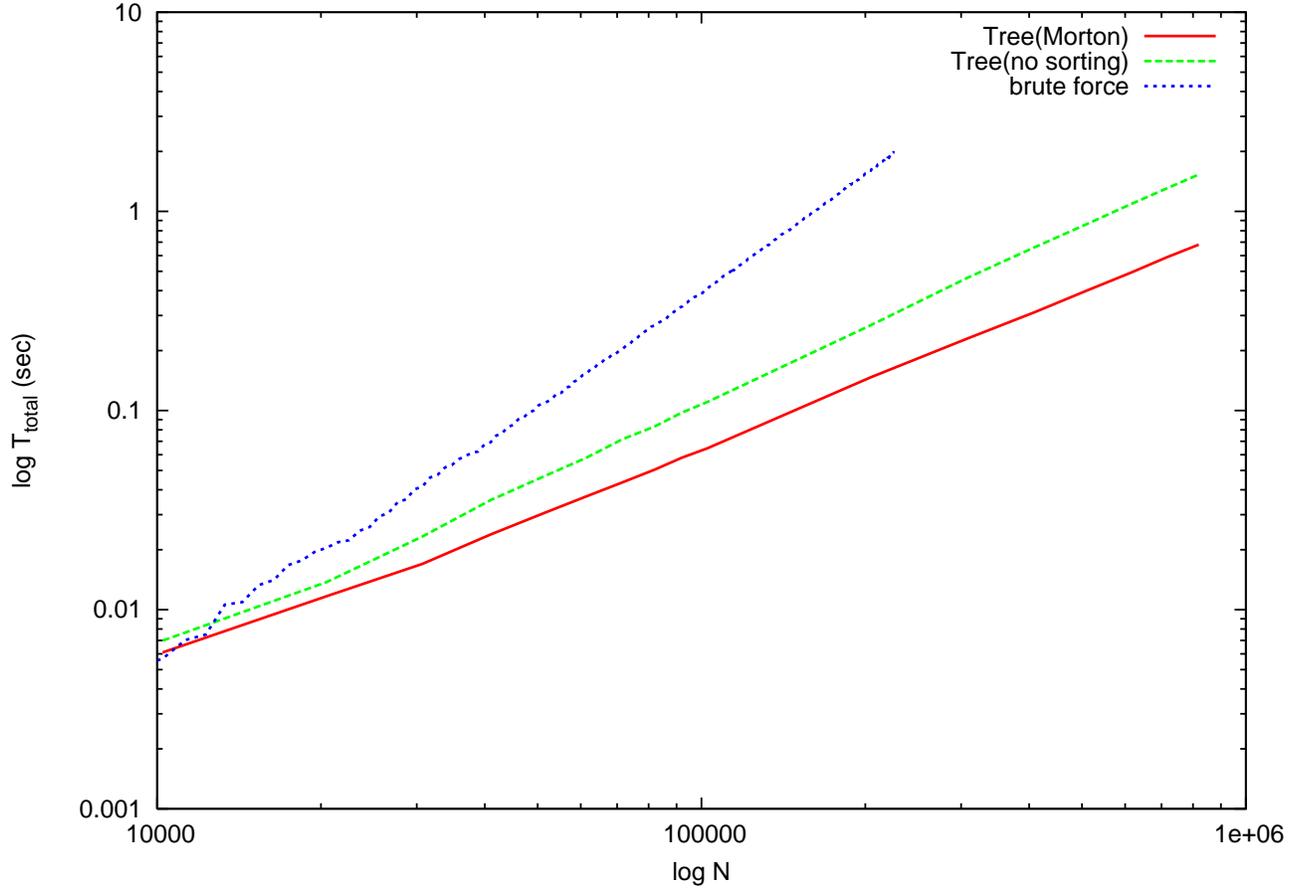}
\caption{Comparison between the oct-tree method on GPU and the brute force method on GPU.
T$_{\rm total}$ as a function of $N$ is plotted for the three cases.
}
\label{time}
\end{figure}

The CAL library offers us a facility to report the cache-hit rate inside the GPU.
In Table \ref{cache}, we show how a cache-hit rate depends on $N$ and an ordering of particles.
This result indicates that the performance of our method on GPU is
significantly affected by an ordering of particles.
In the following, we always use the pre-processing.
Note we could have even better results if we sort particles in
the Peano-Hilbert order, which has been reported to be an optimal order
for locality of data access, as used by some tree codes (e.g., \cite{Warren_1993}). 

In Figure \ref{time}, we present T$_{\rm total}$ as a function of $N$ for three cases:
the oct-tree method with Morton ordering, the oct-tree method without sorting
and the brute force method.
Except for $N < 10,000$, the oct-tree method with the Morton ordering ($\theta = 0.6$)
outperforms the brute force method on GPU.

\begin{table}
\renewcommand{\arraystretch}{1.3}
\caption{Dependence of a computing speed on $N$ : sorting particles in Morton order}
\label{bench2}
\begin{center}
\begin{tabular}{c|c|c}
\hline
$N$ & T$_{\rm total}$ & T$_{\rm construction}$ \\
\hline
50K  & $3.07 \times 10^{-2} $ & $8.3 \times 10^{-3} $ \\
100K & $6.49 \times 10^{-2} $ & $1.7 \times 10^{-2} $ \\
200K & $1.47 \times 10^{-1} $ & $3.6 \times 10^{-2} $ \\
400K & $3.14 \times 10^{-1} $ & $7.4 \times 10^{-2} $ \\
800K & $6.82 \times 10^{-1} $ & $1.5 \times 10^{-1} $ \\
\hline
\end{tabular}
\end{center}
\end{table}

\begin{table}
\renewcommand{\arraystretch}{1.3}
\caption{Dependence of a cache-hit rate on $N$ with a different ordering of particles}
\label{cache}
\begin{center}
\begin{tabular}{c|c|c}
\hline
$N$ & no sorting & Morton order \\
\hline
50K  & 60 \% & 87 \% \\
100K & 51 \% & 82 \% \\
200K & 45 \% & 74 \% \\
400K & 39 \% & 73 \% \\
800K & 35 \% & 69 \% \\
\hline
\end{tabular}
\end{center}
\end{table}

Next, we see how T$_{\rm total}$ depends on $\theta$ that controls error bound
of the oct-tree method.
Larger $\theta$, more distant particles are replaced with the multipole expansion.
In other words, with smaller $\theta$, we require to compute larger number
of force calculations hence it will take longer time.
At the same time, an error due to multipole expansion decreases.
Practically, the force-calculation by the oct-tree method with $\theta < 0.1$
is reduced to almost same as a brute-force computation.
In such regime, effectively we do not have any preference to use the oct-tree method.
In Table \ref{theta}, we show the dependence of T$_{\rm total}$ and a cache-hit rate on $\theta$.
In this test, we used $N = 800K$ particles.
In all tests we presented so far, a clear trend is that 
a computing time seems to be solely determined by a cache-hit rate.
Before we have tested, we have expected branch operations would be a bottleneck
of the compute kernel. 
In reality, all matter is a cache-hit rate on GPU.

\begin{table}
\renewcommand{\arraystretch}{1.3}
\caption{Dependence of T$_{\rm total}$ and a cache-hit rate on $\theta$ for $N$ = 800k}
\label{theta}
\begin{center}
\begin{tabular}{c|c|c}
\hline
$\theta$ & T$_{\rm total}$ & cache-hit rate\\
\hline
0.2 & $ 1.65 \times 10^{1} $ & 23 \% \\
0.3 & $ 5.24 \times 10^{0} $ & 36 \% \\
0.4 & $ 2.24 \times 10^{0} $ & 48 \% \\
0.5 & $ 1.14 \times 10^{0} $ & 59 \% \\
0.6 & $ 6.82 \times 10^{-1}$ & 68 \% \\
0.7 & $ 4.92 \times 10^{-1}$ & 75 \% \\
0.8 & $ 3.98 \times 10^{-1}$ & 80 \% \\
0.9 & $ 3.43 \times 10^{-1}$ & 80 \% \\
1.0 & $ 3.10 \times 10^{-1}$ & 82 \% \\
\hline
\end{tabular}
\end{center}
\end{table}

A final test is to see how a precise data layout on GPU affects a computing speed.
The current version of the CAL has a limitation of array data allocated on GPU
such that the maximum size of one-dimensional array is 8,192 and 
the maximum size of a width and height of two-dimensional array is $8,192 \times 8,192$.
With a one-dimensional array, we can do a simulation of $N = 8,192$ at most
if we do not divide particles into a several arrays.
On the other hand, a two-dimensional array allows us to do a simulation with larger N.
Suppose we do a simulation with $N = 819,200$.
We may use either two-dimensional array of $256 \times 3,200$, 
$512 \times 1,600$, or $1,024 \times 800$ for particle data
(here, we need to allocate an array for $2N$ particles that includes space for cells).
For even larger N, we must use larger width to fit particle data on the two-dimensional array.
In the present work, we restrict to use a power of two for the width to make 
address calculation on the GPU simpler.

A problem we found is that a computing speed of our code changes unexpectedly
depending of layout of the two-dimensional array.
In Table \ref{wsize}, we show this unexpected behavior.
Larger the size of width of an array allocated, 
we obtained a slower computing speed that reflects the lower cache-hit rate
despite in all cases amount of required work and data to be sent to/received from the GPU is identical.
Without detailed information on structure of memory component on the GPU board, 
only we can speculate that possibly precise   layout of the array
on GPU memory affects the cache-hit rate.
As a result, we expect a performance of a simulation that requires larger width
is lower than a performance of smaller N runs.

\begin{table}
\renewcommand{\arraystretch}{1.3}
\caption{Dependence of T$_{\rm total}$ and a cache-hit rate on 
a width of a two-dimensional array allocated on GPU for $N = 819,200$}
\label{wsize}
\begin{center}
\begin{tabular}{c|c|c}
\hline
width & T$_{\rm total}$ & cache-hit rate \\
\hline
256  & $6.82 \times 10^{-1} $ & 69 \% \\
512  & $8.06 \times 10^{-1} $ & 60 \% \\
1024 & $9.46 \times 10^{-1} $ & 51 \% \\
2048 & $1.12 \times 10^{0}  $ & 45 \% \\
4096 & $1.32 \times 10^{0}  $ & 39 \% \\
8192 & $1.34 \times 10^{0}  $ & 35 \% \\
\hline
\end{tabular}
\end{center}
\end{table}

\section{Cosmological Simulation}
As an application of the oct-tree method on GPU, 
we have done an astrophysical many-body simulation using our method.
We report the performance statistics of the simulation in this section.

\begin{figure}[t]
\centering
\includegraphics[width=15cm]{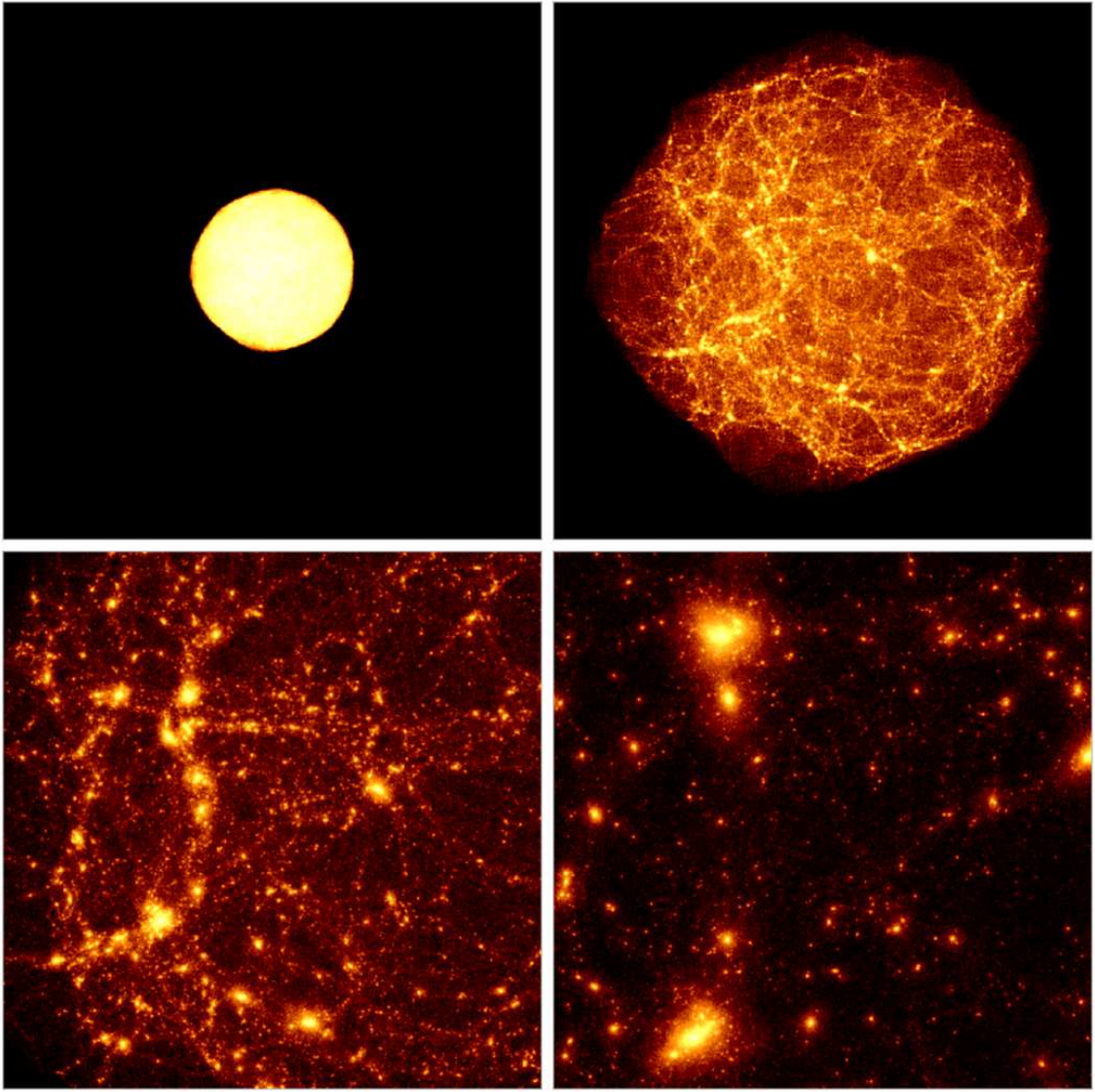}
\caption{Four snapshots of the cosmological simulation. From the top-left to the bottom-right,
each panel shows a snapshot at $z = 8.9, 2.5, 0.93$ and $0$, respectively.
The size of panels is 30 Mpc.}
\label{snap}
\end{figure}

We did a simulation of large-scale structure formation in the universe
with a many-body simulation technique that utilizes the oct-tree method.
We cut out a spherical region of particles from an initial data
that represents initial density fluctuations in the early universe.
We used COSMICS \cite{Bert} to generate the initial data.
The radius of the sphere is 45 mega parse (Mpc) and the number of particles is 2,874,551.
Therefore, each particle represents $5.5 \times 10^9$ solar masses.
We set the initial redshift of the run at $z = 49$ and evolved 
the particles until $z = 0$ (the present time) with 3,584 timesteps.
The total duration in physical unit is $1.4 \times 10^{10}$ years so
that one timestep corresponds to $3.9 \times 10^6$ years.
In this run, we set $\theta = 0.6$ and the softening length ($\epsilon$ in Eq.(\ref{gravity}))
at the present time 4 kpc.
We present a several snapshots of the run in Figure \ref{snap}, 

Our initial condition is not identical but basically same to initial conditions
that have been used in previous entries to Gordon Bell prize
\cite{Kawai_1999, Kawai_2006}.
Precisely, we used larger $N$ and more timesteps than \cite{Kawai_2006}.
We believe such differences do not significantly affect the performance.

\begin{figure}
\centering
\includegraphics[width=7cm,angle=-90]{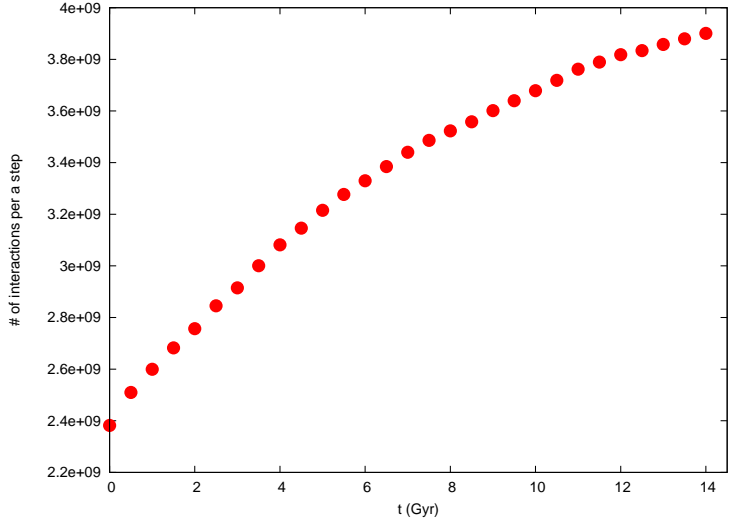}
\caption{A number of force interaction per a step as function of physical time (in Gyr)}
\label{count}
\end{figure}

We measured the wall-clock time reported from the host computer.
This run took 20,860 seconds (5.79 hours) from $z = 49$ to $z = 0$.
This time includes the time required for file I/O and miscellaneous calculations.
During the run, we made the program to dump a snapshot every 128 timesteps.
In total, we have obtained 29 snapshots including an initial snapshot.
We used the snapshots to estimate the total number of interaction counts as follows.
For each snapshot, we separately run our code without utilizing GPU
to count a number of force interactions required 
to calculate gravity for one timestep.
As structure emerged from initial density fluctuations, 
a number of force interactions per a step was increasing as shown in Figure \ref{count}.
Based on this data, we estimated the total number of force interaction
of the GPU run by integrating a function obtained
by connecting points shown in Figure \ref{count} with lines.
The result was $\sim 1.196 \times 10^{13}$ force interactions.
With a conventional floating-point operations count of 38
adopted by previous authors, 
the whole run effectively executed $\sim 4.55 \times 10^{14}$ flops.
Therefore, we have obtained a sustained performance of 21.8 Gflops.
The price per performance is \$41.6/Gflops
that is two and half times better than the most recent result of \$105/Gflops \cite{Kawai_2006}.

\section{Comparisons to Previous Work}
\subsection{Oct-tree Texture on GPU}
In \cite{Lefebvre_2005}, they have implemented oct-tree data structure
for a texture mapping and traversal algorithm of the tree on GPU.
Due to limitations of a GPU and software for GPU programming at that time, 
their method seems to be restricted to applications of computer graphics.
A critical point is that they limit the possible depth of the tree
so that we can not directly employ their implementation for our purpose.

\subsection{Fast Multipole Method on GPU}
In \cite{Gumerov}, they have reported an implementation of a fast multipole method (FMM) on GPU. 
The FMM is a sophisticated algorithm to evaluate long-range force interaction
with computational complexity of $O(N)$.
In the FMM, in addition to the replacement of distant particles with multipole expansions, 
it utilizes local expansions to evaluate force acting on a group of particles.
They have reported for $N = 1,048,576$ it took 0.98 sec with $p = 4$
in their table 9 where $p$ is a parameter that controls error bound.
Their figure 10 indicates that the average relative error obtained with $p = 4$ is $\sim 2 \times 10^{-4}$
that is comparable to the relative error obtained with the oct-tree method with $\theta \sim 0.5 - 0.6$.
Note they have used a randomly distributed particle distribution in a cube.

We did a test with a similar particle distribution for comparison. 
Our code with the size of width 256 took 0.73 sec for $N = 1,048,576$ with $\theta = 0.6$.
The cache-hit rate of the test is 70 \%.
There are many possible explanations to the better performance we obtained
such as (1) the RV770 GPU used in the present work is different from
and newer than their Nvidia G80 GPU and 
(2) even though the FMM is better computational complexity than the tree method,
its algorithmic complexity may cause slower performance on GPU.

Generally, the FMM is well suitable to applications that require long-range force
interaction with uniformly distributed particles/sources
while the oct-tree method is more robust to highly clustered particles 
that typically arises in astrophysical many-body simulations.
Furthermore, the oct-tree method is also efficient to compute general short-range force interaction.
A typical example in astrophysical simulations is the smoothed particle hydrodynamics (SPH) method
\cite{Gingold_1977, Lucy_1977}.
We believe that our method is more suitable than 
\cite{Gumerov} for our purpose of astrophysical applications.

\section{Conclusion}
In this paper, we describe our implementation of the oct-tree method on RV770 GPU.
By transforming a recursive tree traversal into an iterative procedure, 
we show the execution of the tree traversal with the force-calculation
is practical and efficient on GPU.
As an application of our method, we have done a cosmological simulation.
The sustained performance of the simulation is 21.8 Gflops.
The cost per Gflops obtained with the simulation is \$41.6/Gflops that
is two and half times better than a recent result.

In addition, our implementation shows better performance than the recently
reported FMM code on GPU.
We will get further performance gains by fully utilizing four-vector SIMD operations
of TPs and newer GPU with more stream cores.
Moreover, since 10 - 20 \% of T$_{\rm total}$ is spent on the tree construction,
parallelization of this part using multiple cores will be effective to boost the total performance.
Provided that we can easily extend our code to implement
a force-calculation for short-range interaction such the SPH method, 
we believe that a future extended version of our code will enable us 
to do a realistic astrophysical simulation that involves baryon physics
with $N > 1,000,000$ very rapidly.

\section*{Acknowledgment}
The author would like to thank M.~Sato and K.~Fujiwara
for their efforts to utilize RV770 GPU for astrophysical many-body simulations.
A part of this work is based on their undergraduate thesis 2008
at the University of Aizu.

\end{document}